\documentclass[aps,prb,reprint,superscriptaddress,showpacs]{revtex4-1}
\usepackage{graphicx}
\usepackage{bm}
\begin{document}

\title{Pair-density wave states through spin-orbit coupling in multilayer superconductors}

\author{Tomohiro Yoshida}
\affiliation{Department of Physics, Niigata University, Niigata 950-2181, Japan}
\author{Manfred Sigrist}
\affiliation{Theoretische Physik, ETH-Z\"urich, 8093 Z\"urich, Switzerland}
\author{Youichi Yanase}
\affiliation{Department of Physics, Niigata University, Niigata 950-2181, Japan}

\date{\today}

\begin{abstract}
Spin singlet superconductors with quasi-two dimensional multilayer structure are studied in
high magnetic fields. Specifically we concentrate on bi- and tri-layer systems whose layers by symmetry 
are subject Rashba-type spin-orbit coupling. The combination of magnetic field and spin-orbit coupling
leads to a first order phase transition between different states of layer-dependent superconducting order parameters upon rising the magnetic field. 
In this context we distinguish the low-field Bardeen-Cooper-Schrieffer state where all layers have order parameters of the same sign and the high-field 
pair-density wave state where the layer-dependent order parameters change the sign at the center layer. 
We also show that progressive paramagnetic limiting effects yield additional features in the $H$-$T$ phase diagram. As possible 
realizations of such unusual superconducting phases we consider artificial superlattices of 
$\rm{CeCoIn_5}$, as well as some multilayer high-$T_{\rm c}$ cuprates. 
\end{abstract}


\maketitle
\section{INTRODUCTION}
Time reversal and inversion are two of the most important symmetries for the formation of Cooper pairs in a superconductor. Interesting new features of superconductivity arise when one or both of them are missing. One intriguing phase in the absence of time reversal symmetry is the so-called Fulde-Ferrell-Larkin-Ovchinnikov (FFLO) state - a spatially modulated superconducting condensate induced by the spin-splitting of the Fermi surface due to high magnetic fields.~\cite{PhysRev.135.A550,SovPhysJETP.20.762} This phase has been discussed in many contexts, for condensed matter systems~\cite{Nature.425.51,Bianchi_FFLO} and ultra-cold atomic gases,~\cite{Liao} as well as in nuclear physics.~\cite{RevModPhys.76.263} The other interesting class is represented by superconductors without inversion center in the crystal lattice. These so-called non-centrosymmetric superconductors are also characterized by spin-splitting of the Fermi surface, in this case due to anti-symmetric spin-orbit coupling, and form Cooper pairs of mixed parity.~\cite{Springer} Interestingly, both types of situations have been recently discussed in the context of Ce-based heavy Fermion superconductors. A phase observed in CeCoIn$_5$ at low-temperatures and rather high magnetic fields is considered a possible candidate for an FFLO state.~\cite{Nature.425.51,Bianchi_FFLO,JPSJ.76.051005,kumagai2011} Most striking properties of unusually high upper critical fields have been reported for CeIrSi$_3$ and CeRhSi$_3$ which both have a non-centrosymmetric crystal lattice.~\cite{Springer} In our study we would like to consider a combined effect of spin-splitting through magnetic fields and spin-orbit coupling in an artificially structured heavy Fermion superconductor based on CeCoIn$_5$.

\begin{figure}[htbp]
  \includegraphics[width=85mm]{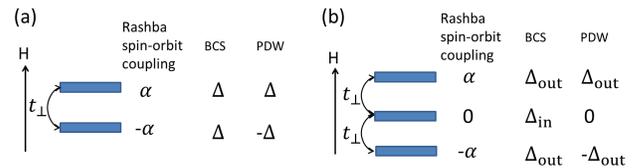}
  \caption{(Color online) Schematic figures of (a) bi- and (b) tri-layer system.
            Thick bars describe the two dimensional conducting planes 
            which are coupled with each other through the inter-layer coupling $t_{\perp}$. 
            The structures of modulated Rashba spin-orbit coupling and 
            superconducting order parameters in BCS and PDW states are  
            shown in the right hand side of each figure. See the text for details. 
    \label{fig:fig1}}
\end{figure}

Mizukami and his coworkers succeeded recently in fabricating superlattices of 
CeCoIn$_5$ and YbCoIn$_5$, the former a heavy Fermion and the latter an ordinary metal.~\cite{Nat.Phys.7.849} Surprisingly this system is superconducting, if the number of CeCoIn$_5$ layers within a unit cell exceeds two. In view of the fact that the upper critical field remains amazingly high despite the low transition temperature,~\cite{Nat.Phys.7.849} the superconductivity in the superlattice is 
robust against paramagnetic depairing effect. 
It was proposed that the superstructure of this material would have locally 
non-centrosymmetric features while having global inversion centers and may 
lead to unusual properties of the superconducting phase.~\cite{Maruyama-Sigrist-Yanase} 
Actually the multilayered structure would induce a spatially dependent Rashba-type of 
spin-orbit coupling~\cite{JPSJ.79.084701} which would influence the spin structure of 
the electronic bands and suppress the paramagnetic depairing effect.~\cite{Maruyama-Sigrist-Yanase} 
Recent measurement of the angular variation of the upper critical field found evidence 
for such a behavior.~\cite{Goh}
In this study, we investigate an unusual form of the superconducting order parameter induced by 
the modulated Rashba spin-orbit coupling in the magnetic field. 
As it will turn out to involve a modulation on the length scale of lattice constant, 
the phase may be viewed as a pair-density wave (PDW) state. 
Here we will restrict to models of bi- and tri-layer systems as depicted schematically in Fig.~\ref{fig:fig1}.

\section{FORMULATION}
For our discussion we analyze the following model of a multilayered system having the inter-layer hopping and Rashba spin-orbit coupling, 
\begin{eqnarray}
H&=&\sum_{{\bm k},s,m}\xi({\bm k})c^\dagger_{{\bm k}sm}c_{{\bm k}sm}
~+t_\perp \sum_{{\bm k},s,\langle m,m'\rangle} c^\dagger_{{\bm k}sm}c_{{\bm k}sm'} \nonumber \\
&&+\sum_{{\bm k},{\bm k}',m}V({\bm k},{\bm k}')c^\dagger_{{\bm k}\uparrow m}c^\dagger_{-{\bm k}\downarrow m}
c_{-{\bm k}'\downarrow m}c_{{\bm k}'\uparrow m} \nonumber\\
&&+\sum_{{\bm k},s,s',m}\alpha_m {\bm g}({\bm k})\cdot {\bm \sigma}_{ss'}c^\dagger_{{\bm k}sm}c_{{\bm k}s'm} 
\nonumber \\
&&-\mu_{\rm B}H\sum_{{\bm k},s,m}sc^\dagger_{{\bm k}sm}c_{{\bm k}sm}, 
\label{eq:eq1}
\end{eqnarray}
where 
$m$ is the index of layers. Note that we restrict here to one set of superconducting layers and ignore the 
normal metal part of the superlattice, assuming that the coupling
between heavy Fermion and normal metal part is small.
This is a reasonable approximation for the superlattice of CeCoIn$_5$, 
since the large mismatch in the Fermi velocities between CeCoIn$_5$ and YbCoIn$_5$ at the interface suppresses the proximity effect.~\cite{She-Balatsky}
For the band structure we assume a simple nearest-neighbor hopping tight binding form 
$\xi({\bm k}) = -2t (\cos k_x + \cos k_y) -\mu$ on a square lattice with the chemical potential $\mu/t = 2$, 
and a small inter-layer hopping $t_\perp/t =0.1 $. The symbol $\langle m,m'\rangle$ denotes the summation over nearest-neighbor layers. It is important to note that the following results do not qualitatively depend on the details of the band structure. We also use an $s$-wave superconducting state by assuming $V({\bm k},{\bm k}')=-V$ (onsite attractive interaction) for simplicity. Although CeCoIn$_5$ is believed to be a $d$-wave superconductor and would have a small odd-parity component admixed with the local non-centrosymmetricity, the features found below 
are independent of the pairing symmetry as long as spin-singlet pairing is predominant.
In fact, we have confirmed that the model for $d$-wave superconductivity gives 
the qualitatively same results (see Appendix). 
Hence, for the sake of numerical accuracy, we restrict ourselves to the simpler $s$-wave state. 
Our choice $V/t=1.7$ gives rise to the critical temperature 
$k_{\rm B}T_{{\rm c}0}/t = 0.0255$ 
for both bi-layer and tri-layer systems with $\alpha=0$, ignoring phase fluctuations. 

For our purpose it is sufficient to choose a $g$-vector with Rashba spin-orbit structure, 
${\bm g}({\bm k})=(-\sin k_y,\sin k_x,0)$ without going into microscopic details of the 
compound.~\cite{JPSJ.77.124711}
The coupling constants $\alpha_m$ are layer dependent and antisymmetric with respect to 
reflection at the center of the multilayer structure. In this way the existence of a global inversion center is guaranteed. For example, $(\alpha_1,\alpha_2)=(\alpha,-\alpha)$ for bi-layers, 
and $(\alpha_1,\alpha_2,\alpha_3)=(\alpha,0,-\alpha)$ for tri-layers (see Fig.~\ref{fig:fig1}). 
Paramagnetic limiting and also eventually the PDW phase enter through the Zeeman coupling term, i.e. the last term in Eq.~(\ref{eq:eq1}).

In our discussion we restrict to magnetic fields along {\it c}-axis 
and analyze the model on the basis of the Bogoliubov-de Gennes equation. 
While we include the layer dependence of order parameter $\Delta_m = 
- \sum_{{\bm k}} V \langle c_{-{\bm k}\downarrow m}c_{{\bm k}\uparrow m} \rangle$, 
we ignore the spatial modulation in the plane, such as the vortex state. This simplification 
is not crucial on a qualitative level, if we assume short coherence lengths and a large Ginzburg-Landau parameter
$ \kappa $, the ratio of coherence length and London penetration depth.  
Although phase fluctuations may play an important role for the critical temperature, 
we focus on the low-temperature and high-magnetic field phase, and ignore this aspect.

\section{SUPERCONDUCTING PHASES}
\subsection{Bi-layer system}
\begin{figure*}[htbp]
  \begin{tabular}{cc}
    \begin{minipage}{0.5\hsize}
      \includegraphics[width=70mm]{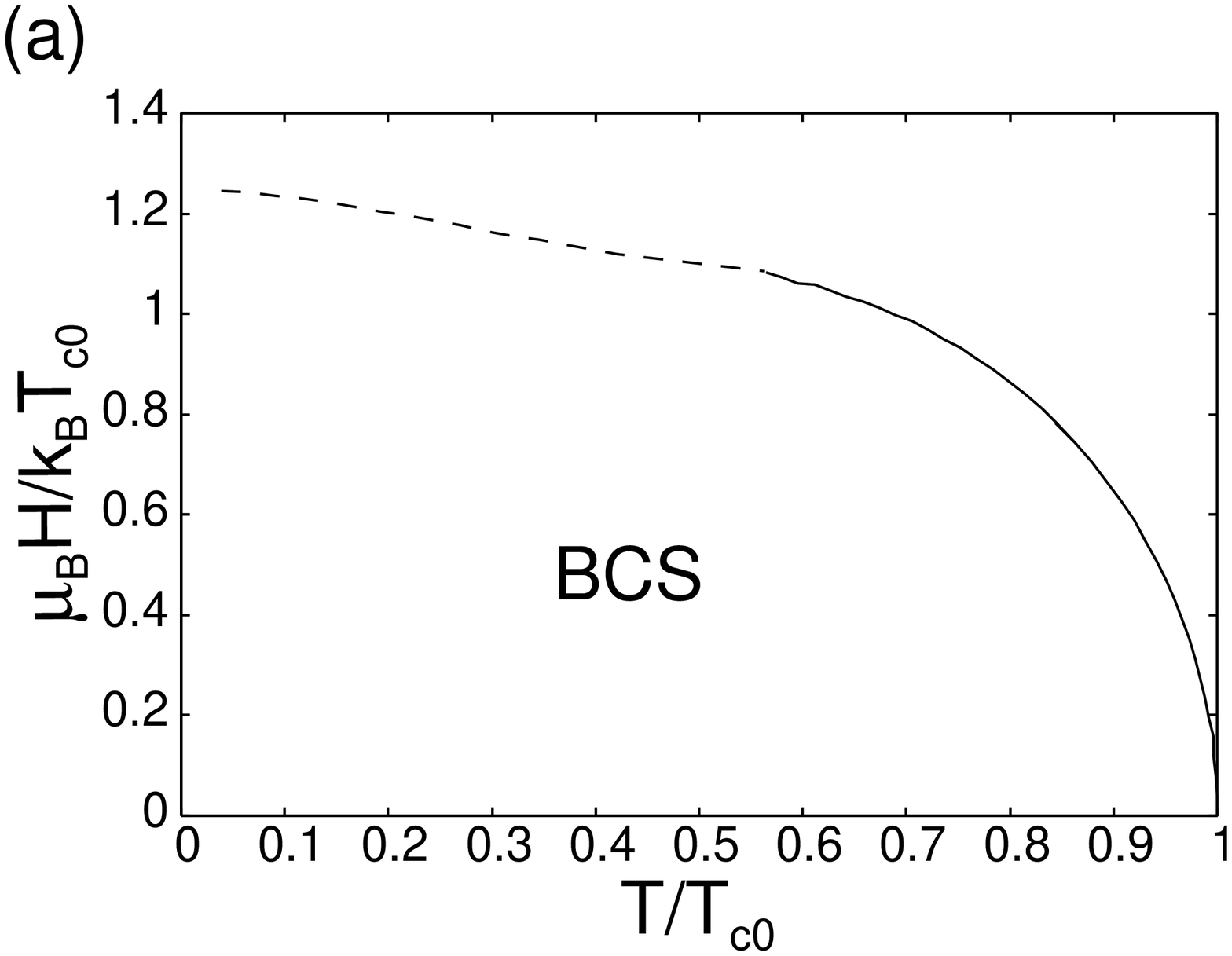}
    \end{minipage}
    \begin{minipage}{0.5\hsize}
      \includegraphics[width=70mm]{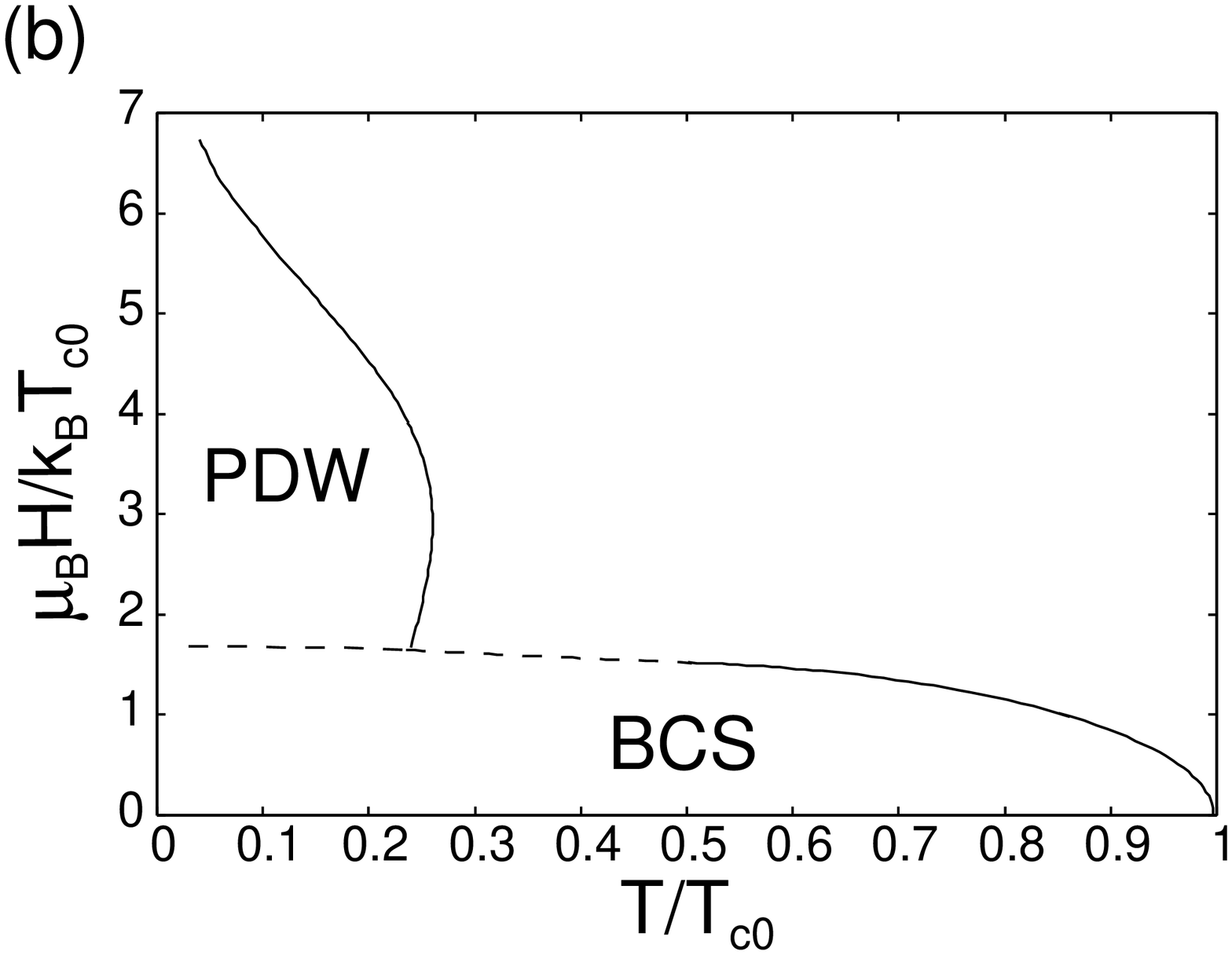}
    \end{minipage}
  \end{tabular}
  \begin{tabular}{cc}
    \begin{minipage}{0.5\hsize}
      \includegraphics[width=70mm]{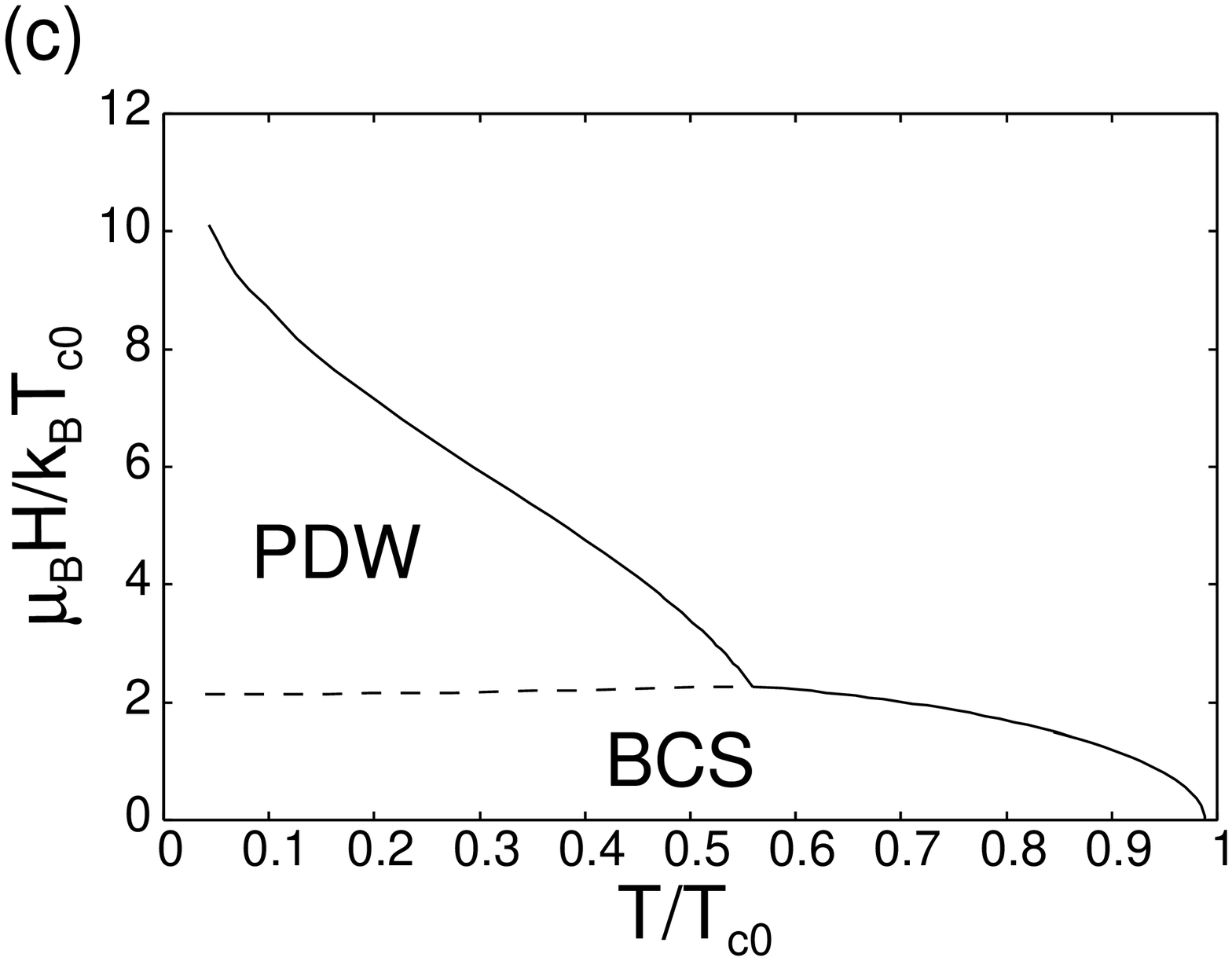}
    \end{minipage}
    \begin{minipage}{0.5\hsize}
      \includegraphics[width=70mm]{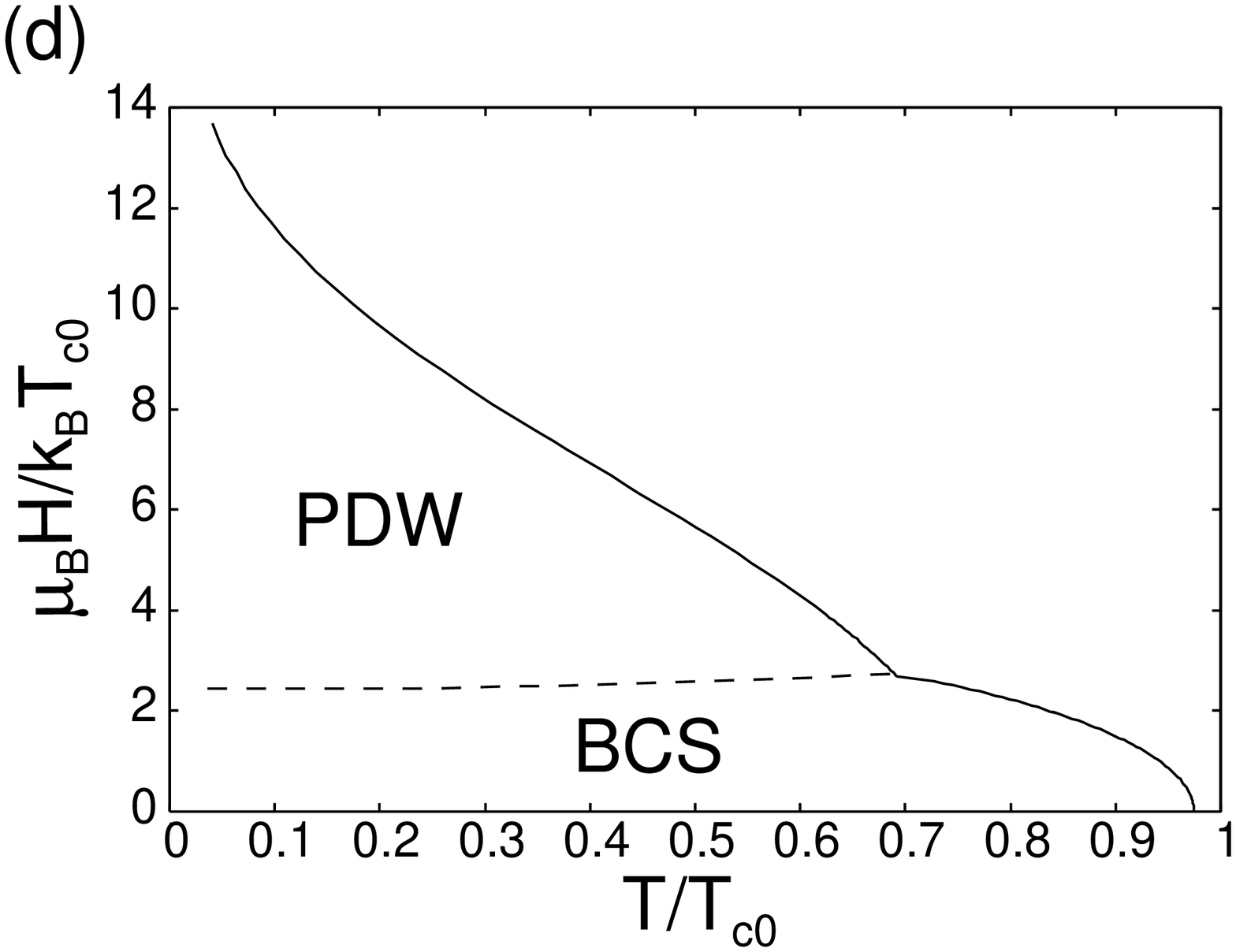}
    \end{minipage}
  \end{tabular}
  \caption{$T$-$H$ phase diagram of bi-layer systems 
for (a) $\alpha/t_\perp=0$, (b) $\alpha/t_\perp=1$, (c) $\alpha/t_\perp=2$, 
and (d) $\alpha/t_\perp=3$. 
The solid and dashed lines show the second- and first-order phase transition lines, 
respectively.
The ``BCS'' and ``PDW'' are explained in the text. 
The temperature $T$ and magnetic field $\mu_{\rm B}H$ in Fig.~\ref{fig:fig2} are 
normalized by the critical temperature $T_{\rm c0}$ at $\alpha/t_\perp=0$. 
    \label{fig:fig2}}
\end{figure*}

First, we address the results for the bi-layer system. 
Figure~\ref{fig:fig2} shows the $T$-$H$ phase diagram for several values of $\alpha/t_\perp$.
For $\alpha/t_\perp=0$ [Fig.~\ref{fig:fig2}(a)], we obtain the conventional phase diagram of a
superconductor subject to paramagnetic limiting.~\cite{JPSJ.76.051005} In this case  
the order parameter is the same in both layers having the same phase,
$(\Delta_1,\Delta_2)=(\Delta,\Delta)$. This state we call Bardeen-Cooper-Schrieffer (BCS) state
is stabilized basically by the inter-layer ``Josephson'' coupling. 
The superconducting phase transition turns first order at low temperatures, 
because of the paramagnetic depairing effect. The in-plane FFLO phase with 
an inhomogeneous order parameter in the two-dimensional plane is supposed to
appear around the first order transition line in Fig.~\ref{fig:fig2}(a). However, in our
approach we ignore the possibility of in-plane modulations as mentioned above.

We now turn to the role of the layer-dependent Rashba spin-orbit coupling. 
Interestingly, the order parameter configuration with a sign change, $(\Delta_1,\Delta_2)=(\Delta,-\Delta)$, 
becomes more stable at high magnetic fields for $\alpha/t_\perp \agt  1$. This state is called 
the pair-density wave (PDW) state in Figs.~\ref{fig:fig2}(b-d),
since the order parameter modulates on length scales of the crystal lattice constant. 
The PDW state was proposed for the $Q$-phase in the bulk 
CeCoIn$_5$,~\cite{PhysRevLett.102.207004,aperis2010} 
but has not been established in microscopic models so far.  
With increasing the spin-orbit coupling $ \alpha $ the PDW state is stabilized in the large parameter range. 
Thus, we here find that the PDW state may be realized in multilayer systems having 
``weak inter-layer coupling'' and ``moderate spin-orbit coupling''. 
We would like to stress that the PDW state is stabilized in the purely superconducting multilayers, 
while a similar inhomogeneous superconducting state can be induced by the proximity effect 
in the ferromagnet-superconductor junction.~\cite{Buzdin}

The stability of the PDW state can be understood on the basis of the spin susceptibility 
and the band structure investigated in Ref.~\onlinecite{Maruyama-Sigrist-Yanase}. 
In the absence of spin-orbit coupling the bi-layer system forms two bands, 
the bonding band and anti-bonding band. Cooper pairs in the PDW state 
would form by inter-band pairing, in contrast to the BCS state which is an intra-band pairing state.
Obviously, the PDW state would not be stable under these circumstances.  
On the other hand, the band structure is described by the two spin-split bands 
on each layer for large $\alpha/t_\perp$. In this case, both BCS and PDW states occur as intra-band Cooper pairing states. 
Although the zero-field transition temperature of the PDW state
is lower than the onset of superconductivity (BCS state), 
a magnetic field along the $c$-axis suppresses the BCS state 
more strongly through paramagnetic depairing. Indeed, the PDW state
has a larger spin susceptibility than the BCS state.~\cite{Maruyama-Sigrist-Yanase} 
Therefore, the PDW state is more robust against paramagnetic limiting than the BCS state,
as shown in Figs.~\ref{fig:fig2}(b-d).

\subsection{Tri-layer system}
\begin{figure}[htbp]
  \includegraphics[width=70mm]{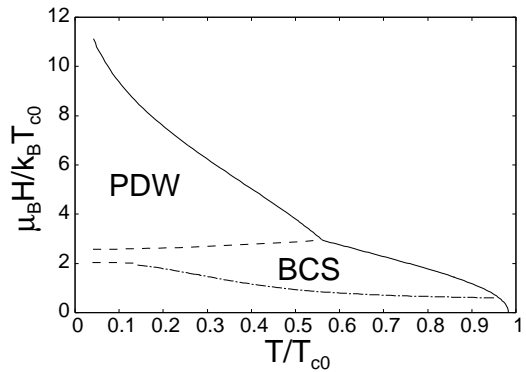}
  \caption{$T$-$H$ phase diagram of tri-layer systems for $\alpha/t_\perp=3$.
    The dash-dotted line shows the crossover field determined by the criterion 
    $\Delta_{\rm in} = \Delta_{\rm out}$. 
    \label{fig:fig3}}
\end{figure}
New aspects can be found for the tri-layer system.
The order parameter has now three components, such that we distinguish
the BCS state as $(\Delta_1,\Delta_2,\Delta_3)=(\Delta_{\rm out},\Delta_{\rm in},\Delta_{\rm out})$ 
and the PDW state as
$(\Delta_1,\Delta_2,\Delta_3)=(\Delta_{\rm out},0,-\Delta_{\rm out})$. 
For $\alpha/t_\perp\alt 2$, the phase diagram is qualitatively the same 
as shown in Fig.~\ref{fig:fig2} for bi-layer systems.  
The range of the PDW state is slightly reduced because of the 
complete suppression of order parameter in the inner layer. 
Increasing spin-orbit coupling to $\alpha/t_\perp=3$ yields an intriguing modification of the 
phase diagram (see Fig.~\ref{fig:fig3}). An further first-order phase transition line
appears within the BCS state, which is absent for bi-layer systems.

\begin{figure}[htbp]
 \includegraphics[width=70mm]{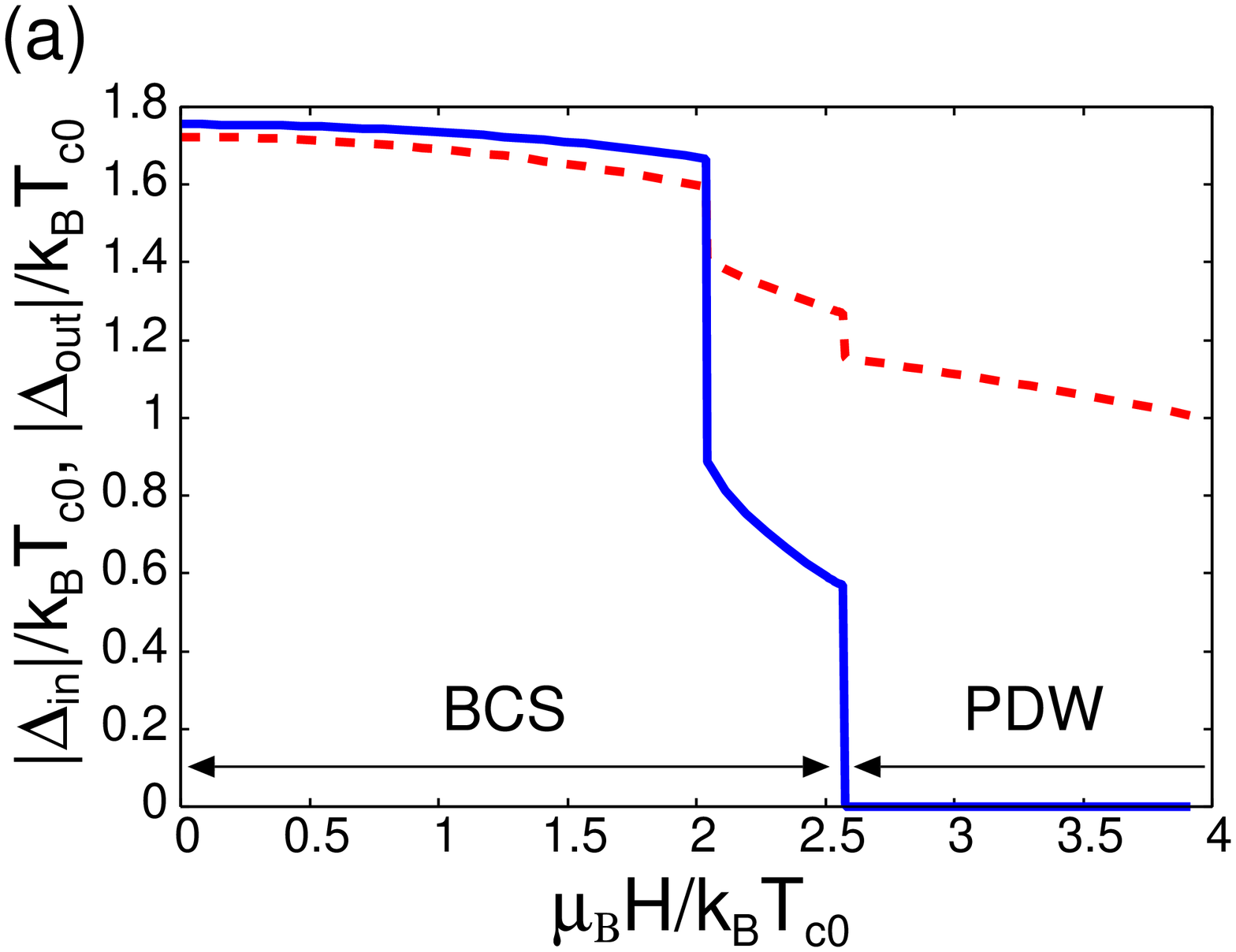}
 \includegraphics[width=70mm]{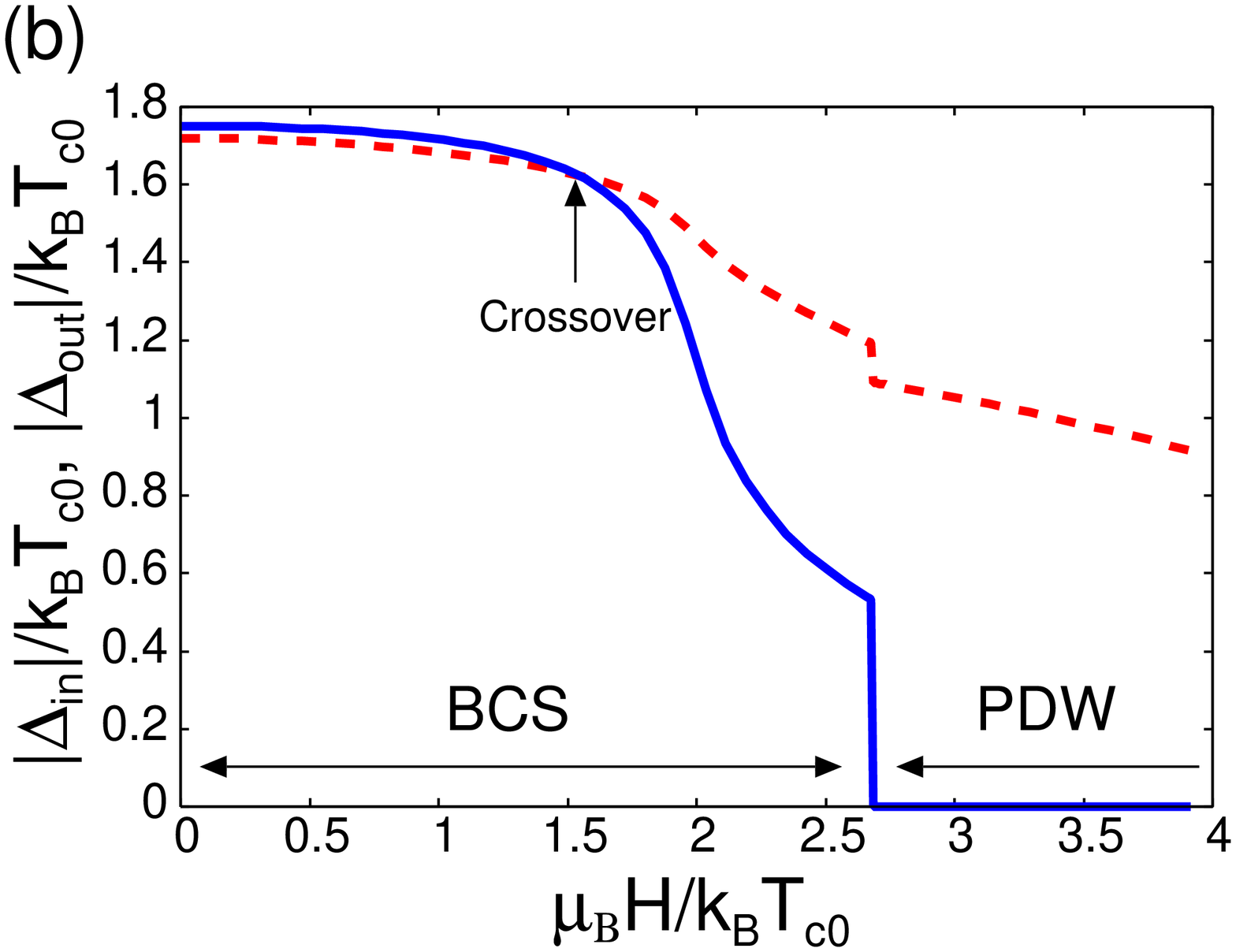}
 \caption{(Color online) Magnetic field dependence of the order parameters 
 in the inner layer $\Delta_{\rm in}$ (solid lines) and outer layers $\Delta_{\rm out}$ (dashed lines) 
 for $\alpha/t_\perp=3$. 
 (a) $T/T_{{\rm c}0}=0.04$ and (b) $T/T_{{\rm c}0}=0.28$. 
 \label{fig:fig4}}
\end{figure}

In order to characterize the first order phase transition in the BCS state, 
we show the magnetic field dependence of the order parameter in Fig.~\ref{fig:fig4}.
At $H=0$, the order parameter of the inner layer $\Delta_{\rm in}$ is slightly larger than 
that of the outer layers $\Delta_{\rm out}$. Upon increasing the magnetic field for $T/T_{{\rm c}0}=0.04$ at $\mu_{\rm B}H/k_{\rm B}T_{{\rm c}0} = 2.04$ an abrupt drop of $\Delta_{\rm in}$ occurs, while $\Delta_{\rm out}$ is only weakly affected [see Fig.~\ref{fig:fig4}(a)]. Indeed, with the layer-dependent spin-orbit coupling the paramagnetic limiting behavior is different for the inner and outer layers. The spin-susceptibility in the superconducting phase is more strongly diminished for the inner than the outer layer when $\alpha/t_\perp > 1$, as the Rashba spin-orbit coupling vanishes by symmetry for the inner layer.~\cite{Maruyama-Sigrist-Yanase} 
The high-field BCS state is stabilized by the more field resistant order parameter component $\Delta_{\rm out}$ of the outer layer. Note that the inter-layer hopping is responsible for the fact that the order parameter components are not behaving independently.  
As the temperature is increased, the abrupt first order phase transition in the BCS state 
turns into a crossover at the critical end point $T/T_{\rm c0} \sim 0.12$. In this regime
the order parameter of the inner layer decreases continuously, as shown in Fig.~\ref{fig:fig4}(b). 

\begin{figure}[htbp]
 \includegraphics[width=70mm]{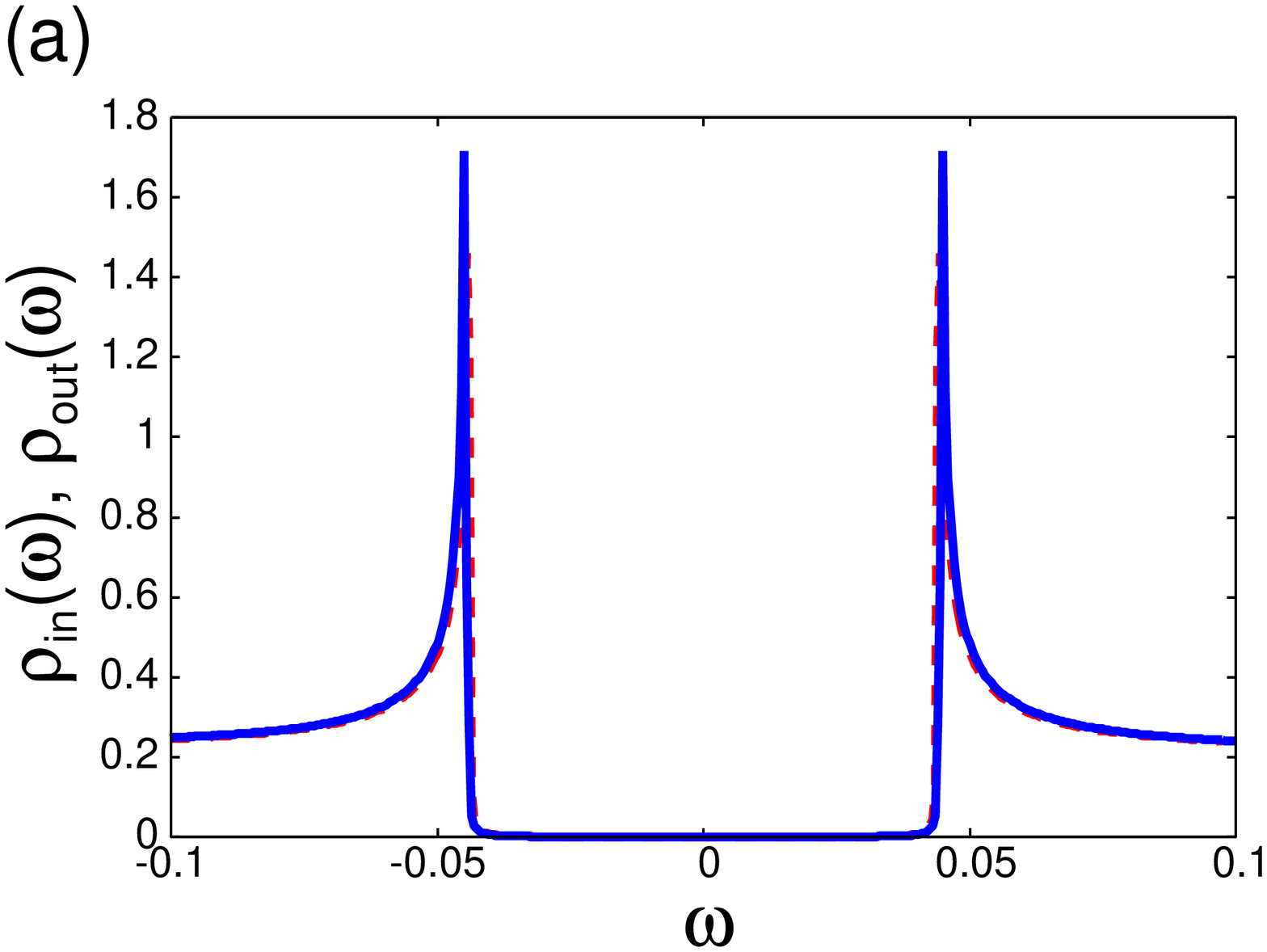}
 \includegraphics[width=70mm]{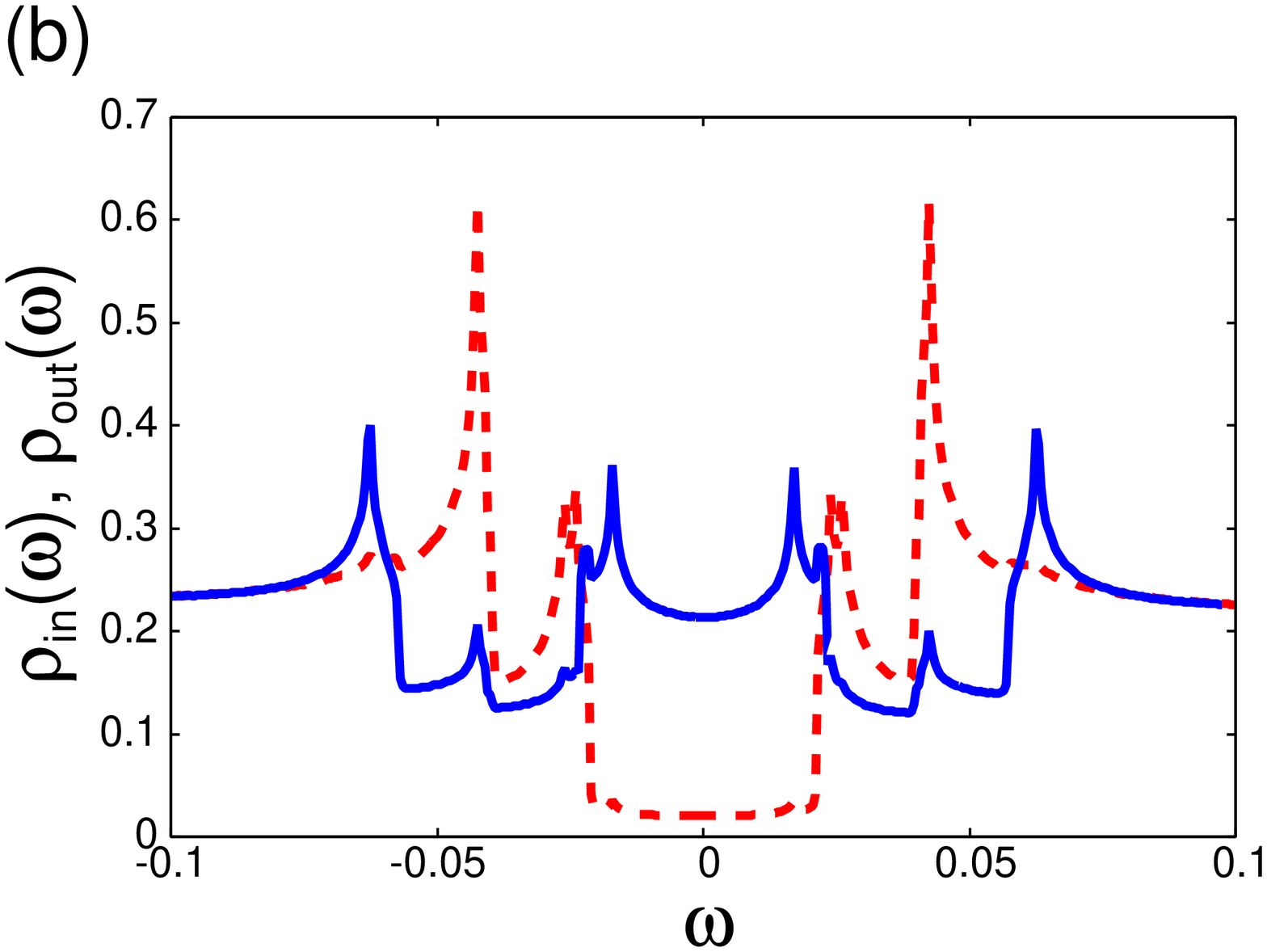}
 \includegraphics[width=70mm]{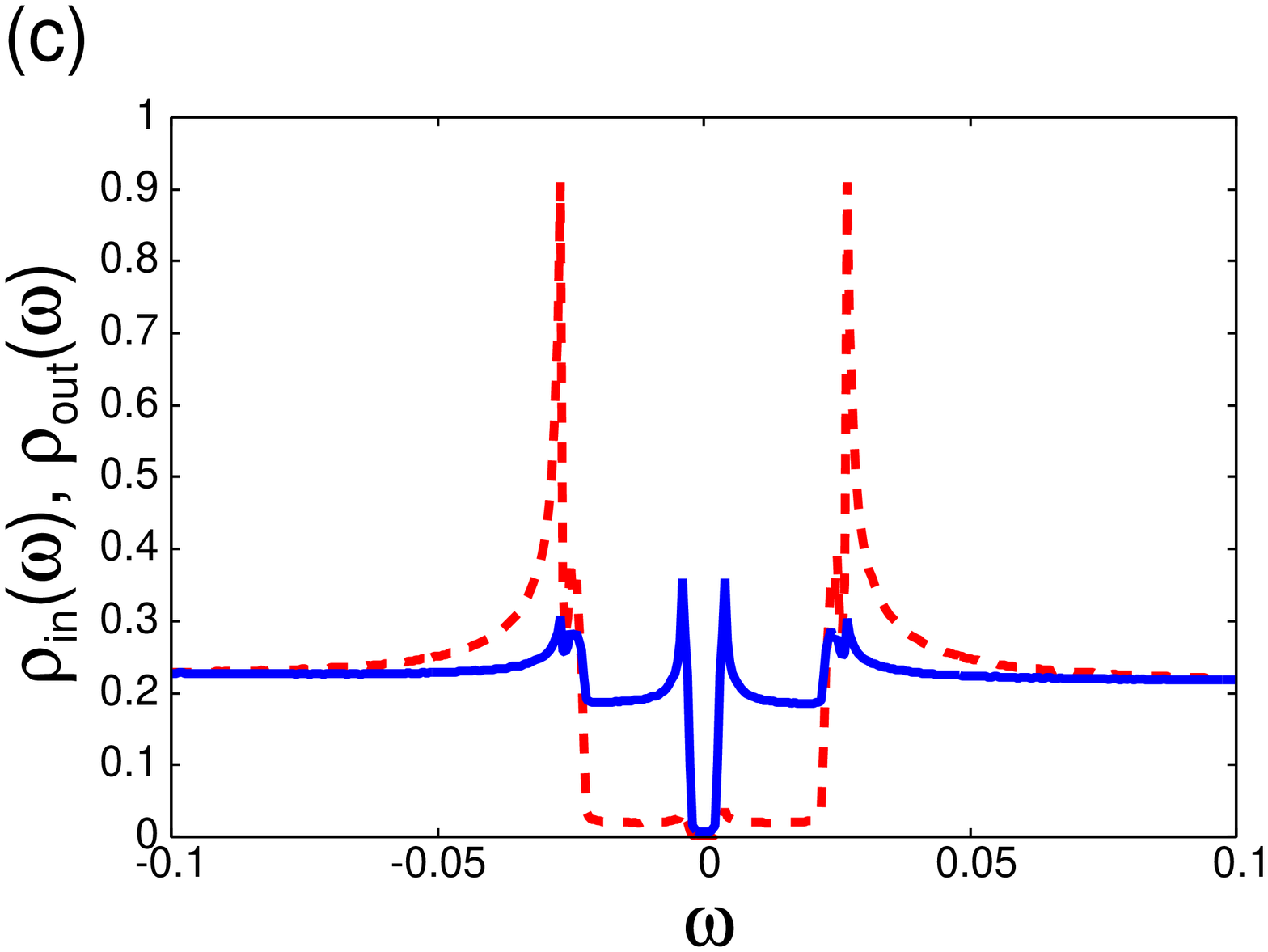}
 \caption{(Color online) DOS in the inner layer $\rho_{\rm in}(\omega)$ (solid lines) 
 and outer layers $\rho_{\rm out}(\omega)$ (dashed lines) at $T/T_{{\rm c}0}=0.04$.  
 (a) $\mu_{\rm B}H/k_{\rm B}T_{{\rm c}0}=0$ (low field BCS state), 
 (b) $\mu_{\rm B}H/k_{\rm B}T_{{\rm c}0}=2.25$ (high field BCS state), 
 and (c) $\mu_{\rm B}H/k_{\rm B}T_{{\rm c}0}=3$ (PDW state), respectively.
 The other parameters are the same as Fig.~\ref{fig:fig3}.
 \label{fig:fig5}}
\end{figure}
The different phases also influence the quasiparticle spectrum which we will discuss here looking at
the density of states (DOS) layer-resolved for the tri-layer system, as depicted in Fig.~\ref{fig:fig5} (solid line for inner and dashed line for outer layer). 
The DOS of the outer layer always shows a gap at the Fermi energy. 
In the absence of a magnetic field the DOS of both the inner and outer layers are 
essentially identical [see Fig.~\ref{fig:fig5}(a)]. Differences appear for finite fields. 
In Fig.~\ref{fig:fig5}(b) we show the DOS in the high-field BCS state, 
which is different for the inner and outer layers. 
The quasiparticle spectrum of the inner layer is gapless, although the order parameter $ \Delta_{\rm in} $ does not disappear. Surprisingly, a small gap on the inner layer opens up again when we enter the PDW state as revealed by Fig.~\ref{fig:fig5}(c), although the order parameter $ \Delta_{\rm in} $ vanishes in this case. The origin of this somewhat counter-intuitive behavior lies in the inter-layer hopping and the energy shifts due to the magnetic field leading to the pair breaking. The inner layer is obviously affected by paramagnetic limiting in the high-field BCS state yielding subgap quasiparticle states. On the other hand, in the PDW state the paramagnetic pair breaking effects almost disappear.~\cite{Maruyama-Sigrist-Yanase}
In this case the intrinsic order parameter of the inner layer vanishes and a small gap appears in 
the quasiparticle excitations because the quasiparticles are extended over all layers 
due to the inter-layer hopping.

\section{SUMARRY AND DISCUSSION}
As mentioned earlier, possible realizations of the situation discussed here could 
be found in artificially grown superlattices of ${\rm CeCoIn_5}$~\cite{Nat.Phys.7.849} 
as well as in multilayer high-$T_{\rm c}$ cuprates.~\cite{Mukuda,Shimizu} 
These multilayer systems are quasi-two-dimensional superconductors and likely 
near the Pauli limit as they are rather robust against orbital depairing due to the very short
coherence lengths.~\cite{Nat.Phys.7.849,Obrien}
Consequently, they satisfy the condition for the phenomenology analyzed in our study. 
Indeed, the reduced paramagnetic limiting due to the Rashba spin-orbit coupling of 
the outer layers has been confirmed in recent experiments for superlattices ${\rm CeCoIn_5}$.~\cite{Goh}
The experiment is approaching to the FFLO state for the field along {\it ab}-plane.~\cite{Goh} 
Our study shows that the PDW state may be stabilized in the magnetic field along {\it c}-axis. 
While this lies in the measurable range for the ``low-$T_{\rm c}$'' system CeCoIn$_5$, 
fields revealing the high-field BCS or PDW state are likely beyond 100 T in high-$T_{\rm c}$ cuprates.    

Finally, we would like to comment on several points of our results. 
First, we restricted to pairing interaction in the spin singlet channel. Spin-singlet pairing is essential 
for the observation of the phase diagram including different BCS states and the PDW state.
If spin triplet pairing compatible with the Rashba-spin-orbit coupling would be dominant, 
paramagnetic limiting would not play a role at all, since the spin susceptibility would keep the
full value of the normal-state Pauli spin susceptibility for magnetic fields along the 
$c$-axis.~\cite{Maruyama-Sigrist-Yanase}
Therefore, no change of phase would be induced by a magnetic field. 
For the intermediate situation, if spin singlet and spin triplet pairing are 
of comparable strength an intriguing mixing of the two channels is possible leading
to a complex phase structure involving spontaneous time reversal symmetry breaking. This
situation will be discussed elsewhere. 
Note, however, that the pairing interactions in the spin triplet channel are expected to be weak 
in the bulk/superlattice of ${\rm CeCoIn_5}$ and high-$T_{\rm c}$ cuprates. 
Even then, weak interactions in the spin triplet channel gives rise to the uniform 
spin triplet Cooper pairs in the PDW state.

Second, we focused on the Pauli-limited superconductor with large Maki parameter, 
neglecting orbital depairing effects, in particular, the aspect of a mixed phase including
flux lines. Although orbital depairing may play quantitatively important roles for the field along {\it c}-axis, 
the PDW state survives moderate orbital depairing effects (moderate Maki parameters) because the spatial modulation of vortex state 
along the {\it ab}-plane is not affecting the $\pi$-phase shift along {\it c}-axis profoundly. 
The situation is quite different for the helical superconducting state in Rashba-type 
non-centrosymmetric superconductors, which is obscured even by the weak orbital 
depairing effect.~\cite{PhysRevB.78.220508}

Third, the PDW state is stabilized not only in the bi- and tri-layers but also in systems with
more layers. In such systems the layers towards the center are subject to paramagnetic limiting
and it is expected that within the BCS state we encounter a progressive demolishing of superconductivity 
in these layers upon increasing field. Eventually, at high enough magnetic field the superconducting
system behaves like the bi-layer system in the PDW state. Due to the fact that these outer layers
are well separated the condition for the appearance of the PDW state is less stringent than in the real bi-layer system. These views have been confirmed by us numerically for the 4-layer system.

In conclusion, we have studied the superconducting state in multilayer systems 
including layer dependent antisymmetric spin-orbit coupling due to the coordination of the layers. 
We found that in a magnetic field along the $c$-axis spin-orbit coupling influences the 
phase of the order parameters of different layers. We find a first order phase transition separating two 
states different by symmetry: the high-field BCS state which is even under reflection at the center of the multilayer system and the PDW state which is odd under reflection. The latter phase 
allows for a considerable enhancement of the upper critical field and generates the uniform order 
parameter for the spin triplet superconductivity. 
This phenomenology 
relies on the fact that orbital depairing is weak, i.e. the coherence length of the superconductor is very short, and that the dominant pairing is in a spin singlet channel. This may apply to the artificially grown multilayer version of CeCoIn$_5$ and some multilayered high-$ T_{\rm c} $ cuprates. 

\begin{acknowledgements}
The authors are grateful to D.F. Agterberg, M. H. Fischer, S. K. Goh, J. Goryo, D. Maruyama, Y. Matsuda, 
T. Shibauchi, and H. Shishido for fruitful discussions. 
This work was supported by KAKENHI (Grant No. 24740230, 23102709, 21102506 and 20740187). 
We also are grateful for financial support of the 
Swiss Nationalfonds, the NCCR MaNEP and the Pauli Center for Theoretical Studies of ETH Zurich. 
\end{acknowledgements}

\appendix*
\section{CASE OF THE D-WAVE SUPERCONDUCTIVITY}
Here we show the phase diagram for $d$-wave superconductivity using a
linearized gap equation.
We define the superconducting susceptibility $\chi^{\rm sc}_{m_1m_2}(q)$
as following, 
\begin{eqnarray}
\chi^{\rm sc}_{m_1m_2}(q)=\int_0^\beta d\tau
 e^{i\Omega_n\tau}\langle B_{{\bm q}m_1}(\tau)B_{{\bm
 q}m_2}(0)\rangle, \nonumber \\
&&
\end{eqnarray}
where $q=({\bm q},i\Omega_n)$,
$\Omega_n=2\pi n/\beta$ with an integer $n$ is a boson Matsubara
frequency, $\beta=1/k_{\rm B}T$, and $B_{{\bm q}m}(\tau)=e^{H\tau}B_{{\bm
q}m}e^{-H\tau}$. $B_{{\bm q}m}=\sum_{\bm
k}c_{{\bm k}+{\bm q}\uparrow m}c_{-{\bm k}\downarrow m}\phi({\bm k})$ is
the creation operator of spin-singlet Cooper pairs with a
center-of-mass momentum ${\bm q}$ on the layer $m$, and $\phi({\bm
k})=\cos k_x-\cos k_y$
shows the momentum dependence of the order parameter with $d$-wave symmetry.
Taking into account the attractive interaction in the $d$-wave channel $V_d$, we obtain the following expression,
\begin{eqnarray}
\hat{\chi}^{\rm sc}(q)=\frac{\hat{\chi}^{\rm sc}_0(q)}{\hat{1}-V_d\hat{\chi}^{\rm
 sc}_0(q)},
\end{eqnarray}  
where $\hat{\chi}^{\rm sc}(q)$ and $\hat{\chi}^{\rm
sc}_0(q)$ are the $M\times M$ matrix representation for the reducible
and irreducible superconducting susceptibility, respectively ($M$ is the number of
layers).
The bare superconducting susceptibility $\hat{\chi}^{\rm
sc}_0(q)$ is obtained as
\begin{widetext}
 \begin{eqnarray}
  \chi^{\rm sc}_{0m_1m_2}(q)&=&\frac{1}{\beta}\sum_{\bm k}\sum_l
   [G_{m_1m_2}^{\uparrow\uparrow}({\bm k}+{\bm q},i\omega_l)G_{m_1m_2}^{\downarrow\downarrow}(-{\bm
   k},i\Omega_n-i\omega_l)\phi({\bm k})\phi^\ast({\bm k}) \nonumber \\
   &&-G_{m_1m_2}^{\uparrow\downarrow}({\bm k}+{\bm
   q},i\omega_l)G_{m_1m_2}^{\downarrow\uparrow}(-{\bm
   k},i\Omega_n-i\omega_l)\phi({\bm k})\phi^\ast(-{\bm k}-{\bm q})],\nonumber \\
&&
\label{eq:eqa2}
 \end{eqnarray}
\end{widetext}
where $\omega_l=\pi(2l+1)/\beta$ is a fermion Matsubara frequency and
$G^{ss'}_{m_1m_2}({\bm k},i\omega_l)$
is the non-interacting Green function.

The superconducting critical temperature $T_{\rm c}$ is determined by the divergence
of the superconducting susceptibility $\hat{\chi}^{\rm sc}({\bf
0},0)$, namely $\hat{1}-V_d\hat{\chi}_0^{\rm sc}({\bm 0},0)=0$
(Thouless criterion). 

\begin{figure*}[htbp]
  \begin{tabular}{cc}
   \begin{minipage}{0.5\hsize}
    \includegraphics[width=70mm]{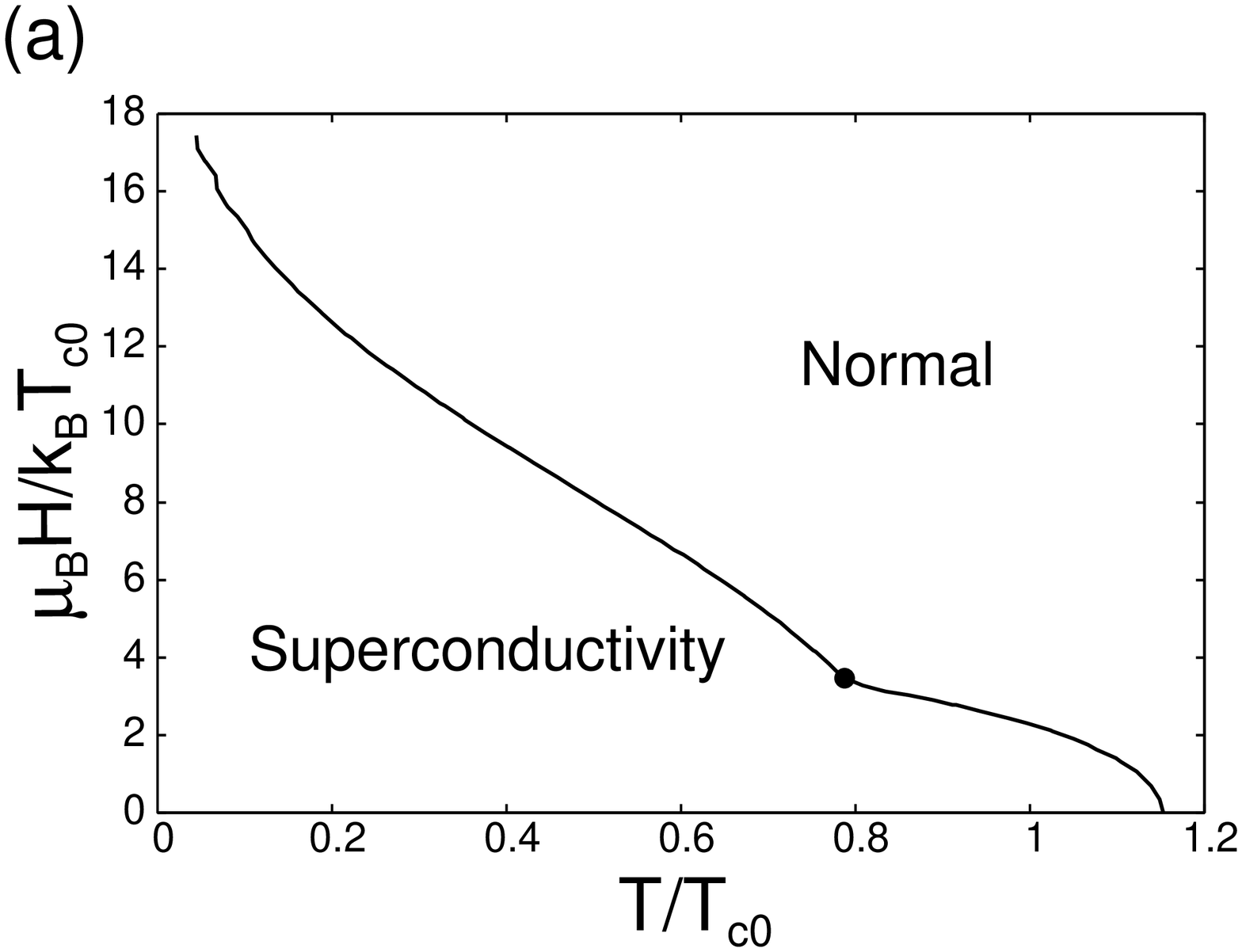}
   \end{minipage}
   \begin{minipage}{0.5\hsize}
    \includegraphics[width=70mm]{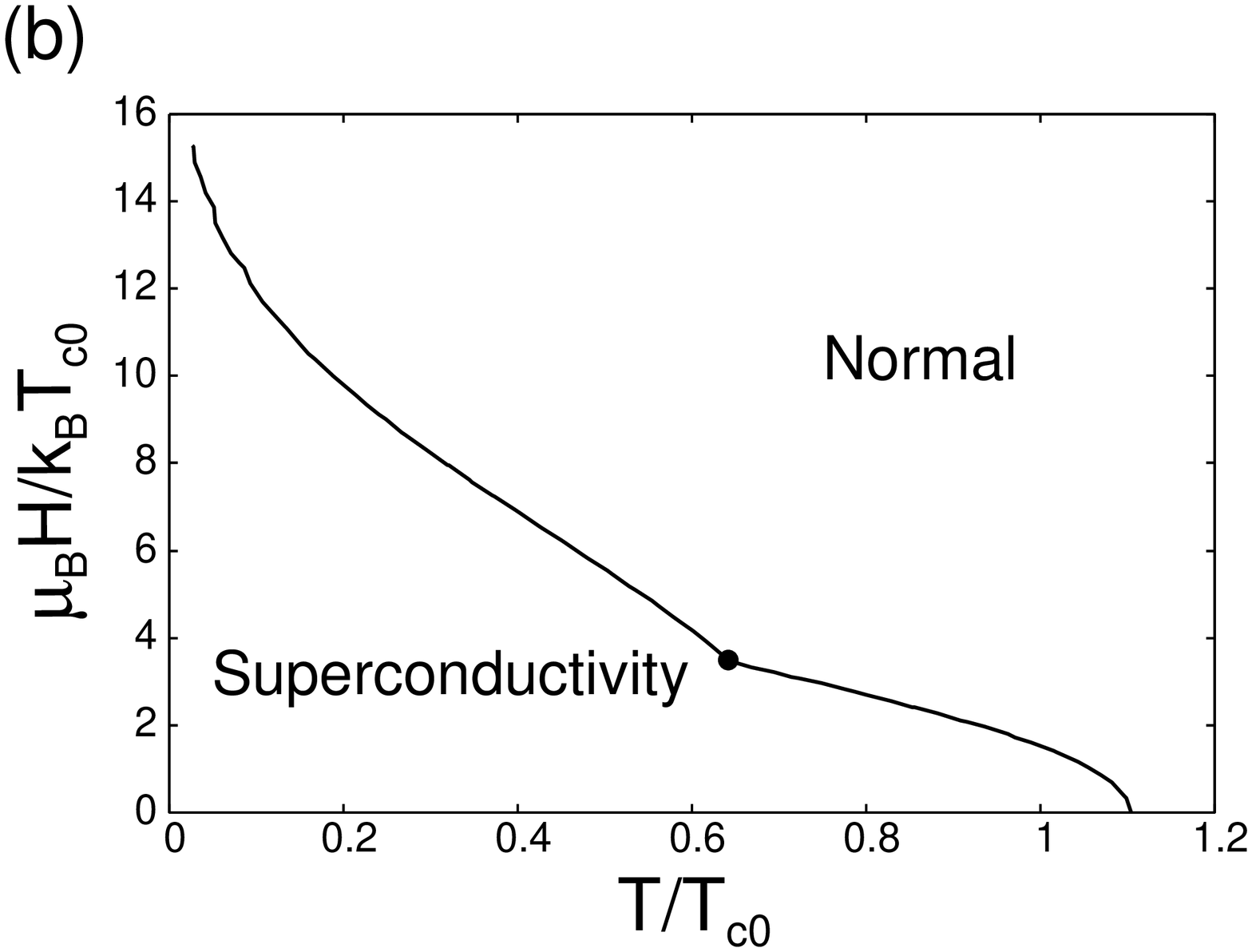}
   \end{minipage}
  \end{tabular}
  \caption{$T$-$H$ phase diagram for (a) bi- and (b) tri-layer system,
  with attractive interaction in the $d$-wave symmetry, $V({\bm k},{\bm k}')=-V_d(\cos k_x-\cos k_y)(\cos k_x'-\cos
  k_y')$. We assume $V_d/t=2.2$ and $\alpha/t_\perp=3$. The critical temperature is
  $k_{\rm B}T_{\rm c0}/t=0.0201$ for bi-layer system and $k_{\rm B}T_{\rm
  c0}/t=0.0202$ for tri-layer system for $\alpha=0$. Other parameters are
  the same as in Fig.~\ref{fig:fig2}.
  The points show the end point of first order transition line between the BCS state to the PDW state.
  \label{fig:fig6}}
 \end{figure*}

In Fig.~\ref{fig:fig6}, we show the phase diagram for bi- and tri-layer
systems.
Although we cannot determine the first order phase transition line
between the BCS and PDW states within the linearized equation, 
the superconducting transition lines in Fig.~\ref{fig:fig6}(a) and
Fig.~\ref{fig:fig6}(b) resemble those in 
Fig.~\ref{fig:fig2}(d) and Fig.~\ref{fig:fig3}, respectively.
Hence, our results in this paper are qualitatively valid for the
$d$-wave superconducting state.

%

\end{document}